\documentclass[aps,
pra,
12pt,
final,
oneside,
onecolumn,
nobibnotes,
nofootinbib,
superscriptaddress,
centertags,
floatfix,
secnumarabic,
notitlepage,
showkeys]{revtex4-2}%

\usepackage{graphicx}
\usepackage{subcaption}
\usepackage{epsfig}
\usepackage{dcolumn}
\usepackage{bm}
\usepackage{amsmath}
\usepackage{amssymb}
\usepackage{xcolor}
\usepackage{nameref}
\usepackage{hyperref}
    \hypersetup{
    colorlinks=true,
    linkcolor=blue,
    filecolor=magenta,      
    urlcolor=cyan,
    pdftitle={Overleaf Example},
    pdfpagemode=FullScreen,
    }
\usepackage[mathlines]{lineno}
\usepackage{comment}

\usepackage[english]{babel}
\begin{document}


\title{Hadronic light-by-light scattering contribution \\to 1S-2S transition in muonium}

\author{V. I. Korobov}
\affiliation{Samara University}
\affiliation{BLTP JINR}
\author{F. A. Martynenko}
\email{f.a.martynenko@gmail.com}
\affiliation{Samara University}
\author{A. P. Martynenko}%
\affiliation{Samara University}%
\author{A. V. Eskin}%
\affiliation{Samara University}%

\date{\today}

\begin{abstract}
We study hadronic light-by-light scattering contribution to the energy interval $1S-2S$ in muonium.
Various amplitudes of interaction of a muon and an electron are constructed, in which the effect of hadronic scattering of 
light-by-light is determined using the transition form factor of two photons into a meson. Their contributions to the particle interaction operator in the case of S-states are obtained in integral form, and to the energy spectrum in numerical form. The contributions of pseudoscalar, scalar, axial vector mesons are taken into account.
\end{abstract}

\maketitle


\section{\label{sec:Introduction} Introduction}

The study of the energy levels of the purely leptonic muonium system has played a central role in many decades of precision calculations with the simplest atoms in quantum electrodynamics
\cite{egs2001}.
The resulting pause in experimental studies of the fine and hyperfine structure of muonium was filled with theoretical calculations of various contributions of the sixth, seventh and higher orders in $\alpha$ \cite{eides1,sgk1}.
As a result, the theoretical accuracy of calculating energy levels in muonium increased significantly and became a benchmark for new experiments. 
It can be said that new experiments on the hyperfine structure in muonium and the 1S-2S energy range are in their final stages.
The latest experimental result on the measurement of the transition frequency 
(1S-2S) in muonium was obtained in the work \cite{mu2000}:
\begin{equation}
\label{f0}
\Delta E(1S-2S)= 2455528941.0 (9.8)~MHz.
\end{equation}

The Muonium Laser Spectroscopy collaboration 
(the Mu-MASS experiment)
aims to measure the (1S-2S)
transition in muonium with a final uncertainty of 10 kHz, providing a 1000-fold improvement on accuracy \cite{crivelli,crivelli1}. 
New result of measurement of the n=2 Lamb shift in muonium
comprises an order of magnitude improvement upon the
previous best measurement \cite{ben}. The MuSEUM (Muonium
Spectroscopy Experiment Using Microwave) collaboration
performed a new precision measurement of the muonium
ground-state hyperfine structure at J-PARC using a high intensity
pulsed muon beam \cite{jparc}.
The MuSEUM collaboration aims to
precisely measure the ground-state hyperfine splitting of
muonium atoms with the accuracy 1 ppb \cite{museum}.
Further increase in the theoretical accuracy of calculations of energy levels in muonium is connected with the calculation of new previously unknown contributions.
Along with purely quantum electrodynamic contributions, there are also contributions due to strong interactions.
Thus, the contribution of hadronic scattering of light-by-light is connected with the production of mesons and cannot be calculated analytically with high accuracy. To calculate it, one can use experimental data on the form factors of the transition of two photons into a meson with certain quantum numbers.
Experimental and theoretical data on the form factors of the transition of two photons into mesons with different spin and parity are constantly updated \cite{tff1,tff2,tff3,tff4}.
The effect of hadronic light-by-light scattering makes an important contribution to the anomalous magnetic moment of the muon \cite{hamm,antoni}.
Other problems in which this contribution was investigated are related to the hyperfine splitting of the ground state of muonium \cite{apm2002,arkasha}
and to the Lamb shift and hyperfine structure in muonic hydrogen \cite{p1,p2,p3,p4}.
The seventh order contributions in $\alpha$, to which this hadronic light-by-light contribution belongs, can have a significant numerical value:
$m\alpha^7=136$ kHz.
But in the case of the hadronic contribution there is an additional factor that leads to a decrease in this value for muonium:
$m^3\alpha^7/F^2_{\pi}\approx 4$ Hz (pseudoscalar meson).
Nevertheless, an exact calculation of such a contribution is of interest for reducing the overall theoretical error.
In our previous works we have investigated the effect of hadronic light-by-light scattering in the Lamb shift and hyperfine structure of muonic hydrogen \cite{apm2018,apm2017}, as well as in the hyperfine structure of muonium, for which contributions from horizontal and vertical exchanges of various mesons were calculated \cite{apm2023}.
Using the previously obtained approaches to calculating this contribution, in this work we performed its calculation for the interval 1S-2S, which should be measured with an accuracy of several kHz.
A large number of works are devoted to the study of the hadronic contribution of light-by-light scattering in solving various problems, among which we would like to note the papers \cite{hamm,narison,pauk}, which contain references to many other publications.

\section{\label{sec:General} General formalism}

\begin{figure}[htbp]
\includegraphics[width=0.97\textwidth]{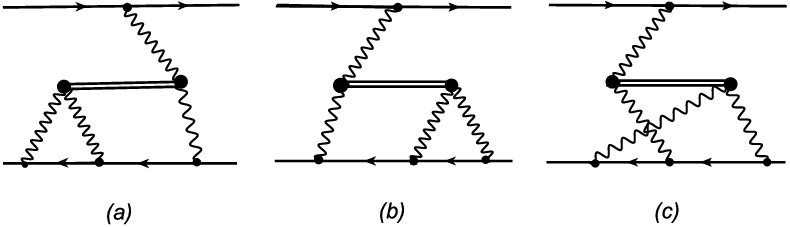}
\caption{Amplitudes of hadronic scattering of light-by-light with three-photon interaction of one lepton. Wavy line corresponds to the virtual photon. The bold dot denotes the form factor of the transition of two photons into a meson.}
\label{ris2}
\end{figure}

In this paper, we study the contribution of the electron-muon interaction amplitudes, which are determined by the effects of light-by-light scattering and are shown in Figs.\ref{ris1},\ref{ris2}.
Fig.\ref{ris1} shows the amplitudes of horizontal and vertical meson exchange. 
In Fig.\ref{ris2}, three of the four photons are attached to the muon line (bottom line on the diagram).
There are three more similar interaction amplitudes in which three photons are attached to an electron line.
When calculating the hyperfine structure in muonium, only the diagrams in Fig.\ref{ris1} are usually taken into account, since the diagrams in Fig.\ref{ris2} can be taken into account if we work with the observed magnetic moment of the muon. But in the case of the energy interval (1S-2S), the amplitudes in Fig.\ref{ris2} also require calculation.

\subsection{The meson interaction amplitudes with horizontal exchange}
\label{subsection1}

\begin{figure}[htbp]
\includegraphics[width=0.97\textwidth]{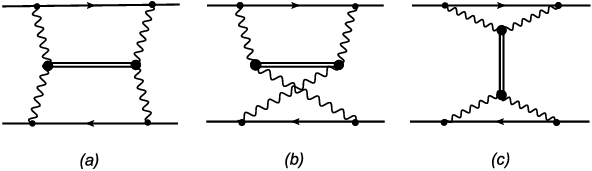}
\caption{Hadronic light-by-light scattering amplitudes with horizontal and vertical exchanges. Wavy line corresponds to the virtual photon. The bold dot denotes the form factor of the transition of two photons into a meson.}
\label{ris1}
\end{figure}

Let us proceed to the calculation of the contributions of various mesons from the horizontal-type amplitudes presented in Fig.\ref{ris1}(a,b).

The first contribution under consideration is determined by pseudoscalar mesons ($\pi^0$, $\eta$, $\eta'$), which are produced as a result of photon-photon interaction.
The interaction vertex of two photons and a pseudoscalar meson is determined by the following expression:
\begin{equation}
\label{f1}
V^{\mu\nu}(k_1,k_2)=i\varepsilon^{\mu\nu\alpha\beta}k_{1\alpha}
k_{2\beta}\frac{\alpha}{\pi F_\pi}F_{\pi^0\gamma\gamma}(k_1^2,k_2^2),
\end{equation}
where the pion decay constant $F_\pi=0.0924$ GeV, $k_1,k_2$ are four momenta of virtual photons. There are various models for the transition form factor that are used to obtain a numerical estimate of the contribution.

The contribution to the Lamb shift (Ls) of the S-energy levels from the horizontal exchange amplitudes with pseudoscalar mesons is determined by the following integral expression in Euclidean space (the index H denotes horizontal exchange):
\begin{equation}
\label{f2}
\Delta E^{Ls,H}_{PS}=-\frac{\alpha^2(Z\alpha)^5\mu^3}{16\pi^4 F_P^2}\int\frac{d k_1 d\Omega_1}{\pi^2}\frac{F_{\pi^0\gamma^\ast\gamma^\ast}(k_1^2,k_2^2)}
{({k_1}^2+a_e^2\cos^2\psi_1)}\times
\end{equation}
\begin{displaymath}
\int\frac{d k_2 d\Omega_2}{\pi^2}
\frac{F_{\pi^0\gamma^\ast\gamma^\ast}(k_1^2,k_2^2)}{(k^2_2+a^2_\mu\cos^2\psi_2)}
\frac{N^{(a+b)}_{PS}}{[(k_1+k_2)^2+\frac{M_P^2}{\Lambda^2}]},
\end{displaymath}
where the expression $N^{(a+b)}_{PS}$ in the numerator is obtained after calculating the trace
as follows:
\begin{equation}
\label{f3}
\varepsilon^{\mu\nu\alpha\beta}k_{1\alpha}k_{2\beta}
\varepsilon^{\lambda\sigma\rho\omega}k_{1\rho}
k_{2\omega}
Tr\Biggl[(\hat q_1+m_1)\gamma^\sigma(\hat p_1-\hat k_1+m_1)\gamma^\mu(\hat p_1+m_1)\frac{1+\gamma^0}{2\sqrt{2}}\hat\varepsilon\times
\end{equation}
\begin{displaymath}
(\hat p_2-m_2)\gamma^\nu
(\hat k_2-p_2+m_2)\gamma^\lambda(\hat q_2-m_2)
\hat\varepsilon\frac{1+\gamma^0}{2\sqrt{2}}\Biggr],
\end{displaymath}
$\varepsilon$ is the polarization vector of state 
with spin 1. The subscript PS denotes the contribution of the pseudoscalar meson.
For a state with spin 0, it is necessary to make a substitution: $\hat\varepsilon\to \gamma_5$.
To pick out the electron-muon states with a certain spin, we use projection operators constructed 
from the 
wave functions of the particles in their rest frame:
\begin{equation}
\hat\Pi_{S=0}=[u(0)\bar v(0)]_{S=0}=\frac{(1+\gamma^0)}{2\sqrt{2}}\gamma_5,~~~
\hat\Pi_{S=1}=[u(0)\bar v(0)]_{S=1}=\frac{(1+\gamma^0)}{2\sqrt{2}}\hat\varepsilon.
\label{f4}
\end{equation}

Using projection operators \eqref{f4}, we extract the contributions to the potential for states with spin 0 
$V(S=0)$ or 1 $V(S=1)$. The contribution to the Lamb shift is then determined using the potential:
\begin{equation}
\label{f5}
V^{Ls}=V(S=0)+\frac{3}{4}V^{hfs},
\end{equation}
where $V^{hfs}$ is the hyperfine part of the potential.
Then the final expression for the numerator 
$N^{(a+b)}_{PS}$ takes the form:
\begin{equation}
\label{f6}
N^{(a+b)}_{PS}=
2a_1 a_2 k_1 k_2\Bigl(
\cos^4\psi_2-\cos^2\psi_2+\cos^4\psi_1-
\cos^2\psi_1-2\cos\Omega\cos\psi_1\cos^3\psi_2-
\end{equation}
\begin{displaymath}
2\cos\Omega\cos\psi^3_1\cos\psi_2+
\cos^2\Omega
\cos^2\psi_1+\cos^2\Omega\cos^2\psi_2+
2\cos^2\Omega\cos^2\psi_1\cos^2\psi_2\Bigr),
\end{displaymath}
which was obtained using the Form package \cite{form}.
In expression \eqref{f6} dimensionless integration momenta are introduced using the $\Lambda$ parameter and $a_1=2m_1/\Lambda$, $a_2=2m_2/\Lambda$.
As usual, to integrate over angles in Euclidean space, one can use the formulas:
\begin{equation}
\label{f7}
d\Omega_1=2\pi \sin^2\psi_1 d\psi_1\sin\theta d\theta,
~~~d\Omega_2=4\pi \sin^2\psi_1 d\psi_1.
\end{equation}

The index $(a + b)$ denotes the contribution of the diagrams $(a)$ and $(b)$ in Fig.~\ref{ris1}.
In obtaining formula \eqref{f2}, the denominators of the propagators are first transformed to the difference of squares in Minkowski space, and then, when moving to Euclidean space, as follows:
\begin{equation}
\label{f8}
\frac{1}{(k_{1,2}^4-4m_{1,2}^2{k_{1,2}^0}^2)}\to \frac{1}{k_{1,2}^2(k_{1,2}^2+4m_{1,2}^2\cos^2\psi_{1,2})}.
\end{equation}

There are various parameterizations for the form factor of the transition of two virtual photons to a pseudoscalar meson \cite{hamm,hamm1}. All of them are based on existing experimental data, which are constantly updated as a result of new measurements. The difference between the parameterizations leads to a change in the result by a value of about one percent, which is of the same order as the numerical accuracy of the calculation of multidimensional integrals. Since our calculations are aimed at obtaining an estimate of this contribution to the shift of S-levels, the question of high accuracy of the calculation is not relevant, and we further use the simplest model of vector dominance (VDM) for the transition form factors
\cite{tff4,apm2002}:
\begin{equation}
\label{MVD}
F_{PS\gamma\gamma}(k_1^2,k_2^2)=\frac{\Lambda_V^4}{(k_1^2-\Lambda_V^2)(k_2^2-\Lambda_V^2)},
\end{equation}
where $\Lambda_V=M_V=M_{\rho}=0.7693~GeV$ is the mass of
$\rho$ meson for the case of $\pi^0$ meson.
Transition form factor data from the BaBar, Belle,
CLEO experiments are obtained mainly for the singly virtual case.

Other pseudoscalar mesons $\eta$, $\eta'$ make contributions determined by the same formula \eqref{f2}. In this case, the decay constants of pseudoscalar mesons $F_\eta$, $F_\eta'$ are determined using the decay width into two photons as follows:
\begin{equation}
\label{gagaga}
F^2_{\eta,\eta'}=\frac{\alpha^2}{64\pi^3}
\frac{M^3_{\eta,\eta'}}{\Gamma(\eta,\eta'\to\gamma\gamma)}.
\end{equation}

Since the form factor \eqref{MVD} is constant at low momenta, the integrals for the contributions \eqref{f2} to the shift (1S-2S) are infrared-finite. The behavior of function \eqref{MVD} at high momenta also ensures the convergence of integrals \eqref{f2} in the ultraviolet range.
The multiple integral in \eqref{f2} is calculated numerically and the result is presented in the Table~\ref{tb1}.
The accuracy of numerical integration is about 1 percent.

\begin{table}[htbp]
\caption{\label{tb1} Hadronic light-by-light contribution to muonium energy interval (1S-2S).
The last three columns in the Table~\ref{tb1} correspond to a horizontal exchange (H), to vertical exchange (V) and
to a lepton three-photon interaction.
}
\bigskip
\begin{tabular}{|c|c|c|c|c|c|c|c|}   \hline
Meson &Mass   & $I^G(J^{PC})$  & $\Lambda$ & $A(M^2,0,0)$  & $\Delta E(H)$ & $\Delta E(V)$   & $\Delta E(3\gamma)$ \\
   & in MeV   &    &   in MeV  &     &   in Hz  &  in Hz   &  in Hz  \\  \hline
$f_1$&1281.9  & $0^+(1^{++})$    & 1040& 0.266  &  0.000002    & 0   &  -0.00009 \\    
    &    &   &     &     $GeV^{-2}$   &  &    &      \\    \hline
$a_1$& 1260 & $1^-(1^{++})$ &1040 & 0.160  &  0.000001 & 0 &-0.00003   \\   
     &    &    &                  &      $GeV^{-2}$        &     &     &           \\    \hline
$f_1$&1426.3 & $0^+(1^{++})$   &926&  0.193  & 0.000001 & 0 &  -0.00003    \\    
    &     &    &                  &      $GeV^{-2}$        &   &     &            \\    \hline
$\sigma$  & 550&$0^+(0^{++})$    & 2000 & -0.596  & -0.00026    & 0.00064  & 0.0676  \\     
    &       &      &                  &     $GeV^{-1}$          &    &   &           \\    \hline
$f_0$  &980 &$0^+(0^{++})$ &2000   &   -0.085  & -0.000001 &  0.000002  &  0.0002    \\     
     &   &     &                  &       $GeV^{-1}$        &    &   &           \\    \hline
$a_0$  &980 &$1^-(0^{++})$   &2000&  -0.086  & -0.000001  & 0.000002 & 0.0002  \\    
    &    &     &                  &       $GeV^{-1}$        &   &   &              \\    \hline
$f_0$  &1370 &$0^+(0^{++})$  &2000&  -0.036  & -0.0000001  &  0.0000002  &  0.00004  \\    
    &     &    &                  &     $GeV^{-1}$   &    &       &                \\    \hline    
$\pi^0$&134.9768 & $1^-(0^{-+})$   &770&  0.025  &  -0.00007   &  0  &   0.0137 \\    
    &   &      &                  &       $GeV^{-1}$   &   &      &                \\    \hline
$\eta$&547.862  & $0^+(0^{-+})$   &774&  0.024  &  -0.000008   &  0 &   0.0019  \\    
    &   &      &                  &        $GeV^{-1}$   &   &      &             \\    \hline    
$\eta'$& 957.78 & $0^+(0^{-+})$   &859&  0.031  & -0.000007  &  0  &   0.0019   \\    
    &   &      &                  &       $GeV^{-1}$   &   &   &               \\    \hline    
Total   & \multicolumn{6}{c}{0.086~Hz}   &      \\  
contribution    & \multicolumn{6}{c}{}    &                            \\  \hline
\end{tabular}
\end{table}

As the second horizontal contribution, we consider the contribution of the axial vector (AV) meson.
The vertex function corresponding to the transition of two photons into a meson has the form \cite{cahn,apm2018}:
\begin{equation}
\label{f9}
T^{\mu\nu}(k_1,k_2)=4\pi i\alpha\varepsilon_{\mu\nu\alpha\beta}(k_1^\alpha k_2^2-k_2^\alpha k_1^2)
\varepsilon_A^\beta A(t^2,k_1^2,k_2^2),
\end{equation}
where $A(t^2,k_1^2,k_2^2)$ is a scalar function of the four-momentum transfer of the virtual photons $k_1$, $k_2$.
The form of the transition form factor $A(t^2,k_1^2,k_2^2)$
of $1^{++}$ meson to two photons was studied experimentally
\cite{L3C,L3Ca,aihara} in different reactions.
In this calculation we use the same parameterization for it as in previous studies of the hadronic contribution in the form (in Euclidean space) \cite{apm2018}:
\begin{equation}
\label{f10}
A(t^2,k_1^2,k_2^2)=A(M_A^2,0,0)F^2_{AV}(k_1^2,k_2^2),~~~F_{AV}(k^2_1,k^2_2)=\frac{\Lambda_A^4}{(\Lambda_A^2+k^2_1)(\Lambda_A^2+k^2_2)},
\end{equation}
where $t=k_1+k_2$ is the four-momentum of the meson.
$A(t^2,k_1^2,k_2^2)$ is a transition form factor which should be found from the experiment.
The effects of off-shellness for exchange by massive AV mesons might be important. 
This effect was investigated in \cite{dorokhov1}, and in \cite{ls} a simple parametrization
was proposed. The simplest way to take it into account is the introduction of the exponential suppression 
factor \cite{ls}:
\begin{equation}
\label{f11}
\frac{A(t^2,0,0)}{A(M_{A}^2,0,0)}
\approx e^{(t^2-M_{A}^2)/M_{A}^2}.
\end{equation}

The numerator of the total amplitude, after the same transformations as in the case of a pseudoscalar meson, can be reduced to the form:
\begin{equation}
\label{f12}
N_{AV}^{(a+b)}=4a_1a_2k_1k_2\Biggl[
k_2^2(\cos^2\psi_1+\cos^2\psi_2+2\cos^3\psi_1\cos\psi_2\cos\Omega-\cos^4\psi_1)+
\end{equation}
\begin{displaymath}
k_1^2(\cos^2\psi_1+\cos^2\psi_2+2\cos^3\psi_2\cos\psi_2\cos\Omega-\cos^4\psi_2)+k_1k_2(-2\cos^2\psi_2\cos\omega-2\cos^2\psi_1\cos\omega-
\end{displaymath}
\begin{displaymath}
\cos\psi_1\cos\psi_2+\cos\psi_1\cos\psi_2\cos^2\Omega+\cos\psi_1\cos^3\psi_2+\cos\psi_2\cos^3\psi_1-4\cos^2\psi_1\cos^2\psi_2\cos\Omega)\Biggr].
\end{displaymath}

Then the total contribution from the amplitudes in 
Fig.~\ref{ris1}(a,b) to the Lamb shift (Ls) of S-levels can be represented by the following integral expression:
\begin{equation}
\label{f13}
\Delta E^{Ls,H}_{AV}=\frac{\alpha^2(Z\alpha)^5\mu^3\Lambda_A^2 A(M_A^2,0,0)^2}{\pi }\int\frac{d k_1 d\Omega_1}{\pi^2}\frac{F^2_{AV}(k_1^2,k_2^2)}{({k_1}^2+a_1^2\cos^2\psi_1)}\times
\end{equation}
\begin{displaymath}
\int\frac{d k_2 d\Omega_2}{\pi^2}
\frac{F^2_{AV}(k_1^2,k_2^2)}{(k^2_2+a^2_2\cos^2\psi_2)}
\frac{N^{(a+b)}_{AV}}{[(k_1+k_2)^2+a_3^2]},
\end{displaymath}
where the index H denotes horizontal exchange.
$a_3=M_A/\Lambda_A$.
The integration in \eqref{f12} is performed numerically 
(see the result in the Table~\ref{tb1}).
We calculated the contribution of three axial vector mesons
$f_1(1285)$, $a_1(1260)$, $f_1(1420)$.

There is another group of scalar (S) mesons that make a significant contribution to the hyperfine splitting in muonium. 
The calculation of the light scalar meson contributions to the anomalous magnetic moment of the muon from hadronic light-by-light scattering was carried out in \cite{narison}.
To calculate their contribution to the interval (1S-2S), we use the general parameterization of the vertex of the interaction of a meson with two virtual photons \cite{pauk,zhou,borisyuk,volkov}:
\begin{equation}
\label{f14}
T_S^{\mu\nu}(t,k_1,k_2)=4\pi\alpha\biggl\{
A(t^2,k_1^2,k_2^2)(g^{\mu\nu}(k_1\cdot k_2)-k_1^\nu k_2^\mu)+
\end{equation}
\begin{displaymath}
B(t^2,k_1^2,k_2^2)(k_2^\mu k_1^2 - k_1^\mu (k_1\cdot k_2))(k_1^\nu k_2^2-k_2^\nu
(k_1\cdot k_2))\biggr\},
\end{displaymath}
where two scalar functions $A(t^2,k_1^2,k_2^2)$, $B(t^2,k_1^2,k_2^2)$ are dependent on three variables, 
$k_{1,2}$ are four momenta of virtual photons, $t$ is the four-momentum of scalar meson.
The term with $A(t^2,k_1^2,k_2^2)$ in \eqref{f14} represents transverse photons interaction, and the
second term represents longitudinal photons interaction.
In the leading order, the contribution of the structure function $A(t^2,k_1^2,k_2^2)$ is decisive.
Four momentum of scalar meson $t$ is equal to $(k_1+k_2)$ for the horizontal exchanges.
The numerator of the sum of the horizontal exchange amplitudes is equal to
\begin{equation}
\label{f15}
N^{(a+b)}_{S}=2a_1a_2k_1k_2\Biggl[-\cos^2\Omega-\cos^2\Omega\cos^2\psi_2-\cos^2\Omega\cos^2\psi_1-\cos^4\psi_1-
\cos^4\psi_2+
\end{equation}
\begin{displaymath}
2\cos\Omega\cos\psi_1\cos\psi_2+
2\cos\Omega\cos\psi_1\cos^3\psi_2+
2\cos\Omega\cos^3\psi_1\cos\psi_2-
\end{displaymath}
\begin{displaymath}
2\cos^2\psi_1\cos^2\psi_2-2\cos^2\Omega\cos^2\psi_1\cos^2\psi_2
\Biggr].
\end{displaymath}

The contribution of scalar meson from two amplitudes in Fig.~\ref{ris1}(a,b) to the bound state energy has 
the following form in Euclidean space:
\begin{equation}
\label{f16}
\Delta E^{Ls,H}_{S}=\frac{\alpha^2(Z\alpha)^5\mu^3}{\pi}\int_0^\infty dk_1\int\frac{ d\Omega_1}{\pi^2}
\int_0^\infty dk_2\int\frac{d\Omega_2}{\pi^2}\frac{A(M_S^2,k_1^2,k_2^2)}{(k_1^2+a_1^2\cos^2\psi_1)}\times
\end{equation}
\begin{displaymath}
\frac{A(M_S^2,k_1^2,k_2^2)}{(k_2^2+a_2^2\cos^2\psi_2)}\frac{N^{(a+b)}_{S}}
{(k_1^2+k_2^2+2k_1k_2\cos\Omega+\frac{M_S^2}{\Lambda_S^2})},
\end{displaymath}
where $M_S$ is the mass of the scalar meson.
For the parameterization of a function $A(M_S^2,k_1^2,k_2^2)$ in the case of scalar meson we use the monopole form for variables $k_1^2$ and $k^2_2$ as in our work \cite{apm3}:
\begin{equation}
\label{f17}
A(M_s^2,k_1^2,k_2^2)=A_S\frac{\Lambda_S^4}{(k_1^2+\Lambda_S^2)(k_2^2+\Lambda_S^2)},
\end{equation}
where numerical values of the $\Lambda$ cutoff parameters are shown in Table~\ref{tb1} for different mesons.
The $S\gamma\gamma$ coupling constant $A_S$ is related to the $S\to\gamma\gamma$ decay rate \cite{apm3,volkov} as follows:
\begin{equation}
\label{f18}
A_S=\sqrt{\frac{4\Gamma_{S\gamma\gamma}}{\pi\alpha^2 M_S^3}},
\end{equation}
where $\Gamma_{S\gamma\gamma}$ is the two photon decay width of the scalar meson.
Although there are no precise experimental data on the two-photon decay widths of scalar mesons yet, it is nevertheless possible to make some estimates of both the parameters $A_S$ and the contributions to the energy spectrum \eqref{f16}, which are listed in Table~\ref{tb1}.
When obtaining all functions in the numerator of the amplitudes \eqref{f2}, \eqref{f13}, \eqref{f16}, we took into account only the leading contributions in $\alpha$, neglecting the transferred momenta of the particles.

\subsection{The meson interaction amplitudes with vertical exchange}
\label{subsection2}

Let us now consider the amplitudes of vertical exchange shown in Fig.~\ref{ris1}(c). As follows from our studies of the muonium hyperfine structure \cite{apm2023}, the contributions of individual vertical mesons are significant.
The contribution from vertical meson exchange is also of seven order in $\alpha$. Its main difference from the contribution of horizontal exchange is that the 4-momentum of the exchange meson is equal to the momentum transferred from the electron to the muon and is therefore of order $\alpha$.

Using the vertex functions from \eqref{f9}, the numerator of such an amplitude can be represented as (see, Fig.~\ref{ris1}(c)):
\begin{equation}
\label{f19}
N^{(c)}_{AV}=\varepsilon_{\mu\nu\alpha\beta}(k_1^\alpha k_2^2-
k_2^\alpha k_1^2)\varepsilon_{\sigma\lambda\rho\omega}(r_1^\rho k_2^2-k_2^\rho k_1^2)\left(g_{\beta\omega}-\frac{t^\beta t^\omega}{M_A^2}\right)
Tr\Biggl[
(\hat q_1+m_1)\gamma^\nu(\hat p_1-\hat k+m_1)\times
\end{equation}
\begin{displaymath}
\gamma^\mu(\hat p_1+m_1)\frac{(1+\gamma^0)}{2\sqrt{2}}\hat\varepsilon (\gamma_5)(\hat p_2-m_2)\gamma^\sigma(\hat r_1-\hat p_2+m_2)
\gamma^\lambda(\hat q_2-m_2)\hat\varepsilon(\gamma_5)
\frac{(1+\gamma^0)}{2\sqrt{2}}\Biggr],
\end{displaymath}
where projection operators onto states with spin 1, 0 \eqref{f4} are introduced, $k_1=k$, $k_2=k-t$, $r_1=r$, $r_2=r-t$. $k$, $r$ are four-momenta in two loops in Fig.~\ref{ris1}(c).
This expression gives zero contribution when constructing operator \eqref{f5}. Thus, the contributions of axial vector mesons occur only in higher order in $\alpha$.

The numerator for the amplitude of vertical (V) exchange with a scalar meson is the following:
\begin{equation}
\label{f20}
N^{(c)}_S=a_1a_2 kr (1+\cos^2\psi_1)(1+\cos^2\psi_2).
\end{equation}

Factorization of integrals with respect to variables k, r takes place. The loop integral with respect to each variable k, r can be calculated analytically in the case of dipole parameterization:
\begin{equation}
\label{21}
I_e=\int \frac{k\sin^2\psi_1(1+\cos^2\psi_1)d\psi_1 dk}{(k^2+a_1^2\cos^2\psi_1)(k^2+1)^2}=\frac{\pi}{4a_1^4\sqrt{1-a_1^2}}
\Biggl[
\sqrt{1-a_1^2}a_1^2+4\sqrt{1-a_1^2}\ln\frac{a_1}{2}+
\end{equation}
\begin{displaymath}
(a_1^2-a_1^4-2)\ln\frac{(2-a_1^2-2\sqrt{1-a_1^2}}{a_1^2}\Biggr].
\end{displaymath}
The integral for the muon loop $I_\mu$ is obtained by replacing $m_1\to m_2$.
Thus, final contribution to the muonium energy spectrum can be represented by the following analytical formula:
\begin{equation}
\label{f22}
\Delta E^{V}_S(nS)=\frac{\alpha^2(Z\alpha)^5\mu^3 A_S^2}{\pi M_S^2n^3}I_eI_\mu,
\end{equation}
neglecting higher order corrections in $\alpha$.
For the numerical evaluation of the contribution to the interval 
(1S-2S) the sigma meson is used (see, Table~\ref{tb1}).

The hadronic contribution of pseudoscalar mesons from vertical exchanges to the S-level shift is also equal to 0 in the leading order $\alpha^7$.
This follows from the general structure of the vertex of the transition of two photons to a pseudoscalar meson
\begin{equation}
\label{f23}
V^{\mu\nu}\sim \varepsilon^{\mu\nu\alpha\beta}k_{1\alpha}k_{2\beta}=0.
\end{equation}
The equality of this expression to zero is obtained when 
$k_1=k$, $k_2=t-k$.

\subsection{
Hadronic amplitudes with three-photon interaction of one lepton}
\label{subsection3}

There is a third class of light-by-light scattering amplitudes with the production of hadrons, shown in Fig.~\ref{ris2}.
In these interaction amplitudes, one lepton (muon or electron) interacts with three virtual photons.
When calculating the hyperfine structure of muonium, these amplitudes are taken into account in the magnetic moment of the electron and muon.
The general approach to calculating such contributions is the same as in the previous paragraphs.

Denoting the loop momenta as $k_1$ and $k_2$, we present general expressions for these amplitudes in the case of pseudoscalar mesons:
\begin{equation}
\label{f24}
T^{(1)}_{PS}=\frac{Z^3\alpha^4}{16\pi^4 F_\pi^2}
\int\frac{d^4k_1}{\pi^2}
\int\frac{d^4k_2}{\pi^2}
\varepsilon^{\mu\nu\omega\rho}t_\omega k_{1\rho}
\varepsilon^{\sigma\lambda\alpha\beta}(-k_{1\alpha}-k_{2\alpha}-
t_{\alpha})k_{2\beta}\times
\end{equation}
\begin{displaymath}
Tr\Biggl\{
\hat\Pi^\ast_{S}
\frac{(\hat q_1+m_1)}
{2m_1}\gamma^\mu\frac{(\hat p_1+m_1)}{2m_1}
\hat\Pi_{S}
\frac{(\hat p_2-m_2)}{2m_2}\gamma^\nu
\frac{(\hat k_1-\hat p_2+m_2)}{(k_1-p_2)^2-m_2^2}\gamma^\sigma\times
\end{displaymath}
\begin{displaymath}
\frac{(\hat q_2+\hat k_2+m_2)}{(q_2+k_2)^2-m_2^2}\gamma^\lambda
\frac{(\hat q_2-m_2)}{2m_2}\Biggr\}\frac{F_{Ps}(0,k_1^2)F_{Ps}((k_1+k_2)^2,k_2^2)}{t^2k_1^2k_2^2(k_1+k_2)^2\left(
k_1^2+\frac{M_{Ps}^2}{\Lambda^2}\right)},
\end{displaymath}
\begin{equation}
\label{f25}
T^{(2)}_{PS}=\frac{Z^3\alpha^4}{16\pi^4 F_\pi^2}
\int\frac{d^4k_1}{\pi^2}
\int\frac{d^4k_2}{\pi^2}
\varepsilon^{\mu\nu\omega\rho}t_\omega (k_{1\rho}-k_{2\rho})
\varepsilon^{\sigma\lambda\alpha\beta}(-k_{1\alpha}-
t_{\alpha})(-k_{2\beta})\times
\end{equation}
\begin{displaymath}
Tr\Biggl\{
\hat\Pi^\ast_{S}\frac{(\hat q_1+m_1)}
{2m_1}\gamma^\mu\frac{(\hat p_1+m_1)}{2m_1}
\hat\Pi_{S}\frac{(\hat p_2-m_2)}{2m_2}\gamma^\sigma
\frac{(\hat k_1+\hat q_2+m_2)}{(k_1+q_2)^2-m_2^2}\gamma^\nu\times
\end{displaymath}
\begin{displaymath}
\frac{(\hat q_2-\hat k_2+m_2)}{(q_2-k_2)^2-m_2^2}\gamma^\lambda
\frac{(\hat q_2-m_2)}{2m_2}\Biggr\}\frac{F_{Ps}(0,k_1^2)F_{Ps}((k_1+k_2)^2,k_2^2)}{t^2k_1^2k_2^2(k_1+k_2)^2\left(
(k_1+k_2)^2+\frac{M_{Ps}^2}{\Lambda^2}\right)},
\end{displaymath}

\begin{equation}
\label{f26}
T^{(3)}_{PS}=\frac{Z^3\alpha^4}{16\pi^4 F_\pi^2}
\int\frac{d^4k_1}{\pi^2}
\int\frac{d^4k_2}{\pi^2}
\varepsilon^{\mu\nu\omega\rho}t_\omega k_{1\rho}
\varepsilon^{\sigma\lambda\alpha\beta}k_{2\alpha}(-k_{1\beta}-k_{2\beta}-t_\beta)\times
\end{equation}
\begin{displaymath}
Tr\Biggl\{
\hat\Pi^\ast_{S}\frac{(\hat q_1+m_1)}
{2m_1}\gamma^\mu\frac{(\hat p_1+m_1)}{2m_1}\hat\Pi_{S}\frac{(\hat p_2-m_2)}{2m_2}\gamma^\sigma
\frac{(\hat k_2-\hat p_2+m_2)}{(k_2-p_2)^2-m_2^2}\gamma^\sigma\times
\end{displaymath}
\begin{displaymath}
\frac{(-\hat q_2-\hat k_1+m_2)}{(q_2+k_2)^2-m_2^2}\gamma^\nu
\frac{(\hat q_2-m_2)}{2m_2}\Biggr\}\frac{F_{Ps}(0,k_1^2)F_{Ps}((k_1+k_2)^2,k_2^2)}{t^2k_1^2k_2^2(k_1+k_2)^2\left(
k_1^2+\frac{M_{Ps}^2}{\Lambda^2}\right)},
\end{displaymath}
where for convenience the charge of the second particle (muon) is designated by Z. 
The projection operators $\hat\Pi_S$ are defined by formulas \eqref{f4}.
The further transformation of the amplitudes is connected with the transition to Euclidean space. Thus, the transformation of the denominators of the propagators of the second particle is determined as follows:
\begin{equation}
\label{f27}
\frac{1}{(k_1^2+2m_2 k^0_1)(k_2^2-2m_2k^0_2)}=
\frac{(k^2_1-2m_2k_1^0)(k^2_2+2m_2k_2^0)}
{(k_1^4-4m_2^2{k^0_1}^2)(k_2^4-4m_2^2{k^0_2}^2)}\to
\end{equation}
\begin{displaymath}
\to
\frac{(k_1^2+2Im_2k_1^0)(k_2^2-2Im_2k_2^0)}{k_1^2k_2^2(k_1^2+4m_2\cos^2\psi_1)(k_2^2+4m_2\cos^2\psi_2)}
\end{displaymath}

After such transformations we obtain the following expressions in the numerator of the amplitudes \eqref{f24},\eqref{f25}, \eqref{f26}:
\begin{equation}
\label{f28}
N^{(1)}_{PS}=
k_1^3k_2^3 \Bigl(2 -\frac{2}{3}\sin^2\psi_2 -\frac{2}{3} \sin^2\psi_1 -2\cos^2\psi_1 + \frac{2}{3} \cos^2\psi_1
\sin^2\psi_2 + \frac{2}{3}\cos^2\psi_1 \cos^2\psi_2 + 
\end{equation}
\begin{displaymath}
\frac{2}{3}\cos\Omega\cos\psi_1\cos\psi_2 - 
\frac{2}{3}\cos\Omega\cos\psi_1\cos\psi_2\sin^2\psi_1 -
\frac{2}{3}\cos\Omega\cos^3\psi_1\cos\psi_2 + 
2/3\cos^2\Omega - 
\end{displaymath}
\begin{displaymath}
\frac{4}{3}\cos\Omega^2\cos\psi_1^2\Bigr) + 
 k_1^4k_2^2\Bigl(\frac{7}{3}\cos\psi_1\cos\psi_2 - \cos\psi_1\cos\psi_2\sin^2\psi_1 - 
 \cos^3\psi_1\cos\psi_2 - 
 \frac{1}{3}\cos\Omega + 
\end{displaymath}
\begin{displaymath}
 \frac{1}{3}\cos\Omega\sin^2\psi_1 - 2\cos\Omega\cos^2\psi_2 + 
\frac{2}{3}\cos\Omega\cos^2\psi_2\sin^2\psi_1 -
\cos\Omega\cos^2\psi_1+ 
\frac{2}{3}\cos\omega\cos\psi_1^2\cos\psi_2^2+ 
\end{displaymath}
\begin{displaymath}
\frac{4}{3}\cos^2\Omega\cos\psi_1\cos\psi_2 \Bigr) +
a_2^2 k_1^2 k_2^2\Bigl( 2\cos\psi_1\cos\psi_2 - \frac{2}{3}\cos\psi_1\cos\psi_2\sin^2\psi_2- 
 \end{displaymath}
\begin{displaymath}
\frac{2}{3}\cos\psi_1
\cos\psi_2\sin^2\psi_1 - \frac{4}{3}\cos^3\psi_1\cos\psi_2 + 
\frac{2}{3}\cos^3\psi_1\cos\psi_2\sin^2\psi_2 + \frac{2}{3}\cos^3\psi_1\cos^3\psi_2 - 
\end{displaymath}
\begin{displaymath}
\frac{2}{3}\cos\Omega\cos\psi_1^2 - \frac{4}{3}\cos\Omega\cos\psi_1^2\cos\psi_2^2 + 
\frac{2}{3}\cos\Omega^2\cos\psi_1\cos\psi_2 \Bigr) + 
\end{displaymath}
\begin{displaymath}
a_2^2k_1^3k_2\Bigl(\frac{5}{3}\cos\psi_1^2\cos\psi_2^2 - \cos\psi_1^2\cos\psi_2^2\sin\psi_1^2 - 
 \cos\psi_1^4\cos\psi_2^2 + 
 \end{displaymath}
\begin{displaymath}
 \frac{1}{3}\cos\Omega\cos\psi_1\cos\psi_2 + 
\frac{1}{3}\cos\Omega\cos\psi_1\cos\psi_2\sin\psi_1^2 - 
 \cos\Omega\cos\psi_1^3\cos\psi_2 \Bigr),
\end{displaymath}
\begin{equation}
\label{f29}
N^{(2)}_{PS}=
 a_2^2 k_1^2 k_2^2 \Bigl(-\frac{2}{3} \cos\Omega^2 \cos\psi_1 \cos\psi_2+
 \frac{8}{3} \cos\Omega \cos^2\psi_1 \cos^2\psi_2-
\frac{2}{3} \cos\Omega \cos^2\psi_1-
\end{equation}
\begin{displaymath}
2 \cos\Omega \cos^2\psi_2+2 \cos^3\psi_1 \cos^3\psi_2-
\frac{16}{3} \cos^3\psi_1 \cos\psi_2-
\cos\psi_1 \cos^3\psi_2+5 \cos\psi_1 \cos\psi_2+
\end{displaymath}
\begin{displaymath}
\frac{5}{3} \cos^3\psi_1 \sin^2\psi_2 \cos\psi_2+
\frac{1}{3} \sin^2\psi_1 \cos\psi_1 \cos^3\psi_2- \sin^2\psi_1 \cos\psi_1 \cos\psi_2-
\end{displaymath}
\begin{displaymath}
\frac{4}{3} \cos\psi_1 \sin^2\psi_2 \cos\psi_2\Bigr)+
a_2^2 k_1^3 k_2 \Bigl(- \cos\Omega \cos^3\psi_1 \cos\psi_2+\frac{1}{3} \cos\Omega \cos\psi_1 \cos\psi_2+
\end{displaymath}
\begin{displaymath}
\frac{1}{3} \cos\Omega \sin^2\psi_1 \cos\psi_1 \cos\psi_2+\frac{11}{3} \cos^2\psi_1 \cos^2\psi_2-\cos^4\psi_1 \cos^2\psi_2-
\sin^2\psi_1 \cos^2\psi_1 \cos^2\psi_2+
\end{displaymath}
\begin{displaymath}
\frac{2}{3} \sin^2\psi_1 \cos^2\psi_2-2 \cos^2\psi_2\Bigr)+
a_2^2 k_1 k_2^3 (-\frac{1}{3} \cos\Omega \cos\psi_1 \cos^3\psi_2+\frac{7}{3} \cos\Omega \cos\psi_1 \cos\psi_2-
\end{displaymath}
\begin{displaymath}
\frac{1}{3} \cos\Omega \cos\psi_1 \sin^2\psi_2 \cos\psi_2-\frac{4}{3} \cos^2\psi_1 \cos^2\psi_2+\frac{1}{3} \cos^2\psi_1 \cos^4\psi_2+
\frac{1}{3} \cos^2\psi_1 \sin^2\psi_2 \cos^2\psi_2+
\end{displaymath}
\begin{displaymath}
\frac{1}{3} \cos^2\psi_1 \sin^2\psi_2- \cos^2\psi_1+\cos^4\psi_2-
\cos^2\psi_2+\frac{1}{3} \sin^2\psi_2 \cos^2\psi_2)+
\end{displaymath}
\begin{displaymath}
a_2^2 k_2^4 \Bigl(-\cos\psi_1 \cos^3\psi_2+ \cos\psi_1 \cos\psi_2-
\frac{1}{3} \cos\psi_1 \sin^2\psi_2 \cos\psi_2\Bigr)+k_1^3 k_2^3 
(-\frac{2}{3} \cos^2\Omega+
\end{displaymath}
\begin{displaymath}
\frac{8}{3} \cos\Omega \cos\psi_1 \cos\psi_2+
2 \cos^2\psi_1 \cos^2\psi_2+\frac{1}{3} \sin^2\psi_1 \cos^2\psi_2+
\frac{5}{3} \cos^2\psi_1 \sin^2\psi_2-\frac{1}{3}\sin^2\psi_1-
\end{displaymath}
\begin{displaymath}
4 \cos^2\psi_1-\sin^2\psi_2-2 \cos^2\psi_2+2)+k_1^4 k_2^2 
(\frac{1}{3} \cos\Omega \sin^2\psi_1-\cos\Omega \cos^2\psi_1-
\frac{1}{3}\cos\Omega-
\end{displaymath}
\begin{displaymath}
\cos^3\psi_1 \cos\psi_2+\frac{7}{3} \cos\psi_1 \cos\psi_2-\sin^2\psi_1 \cos\psi_1 \cos\psi_2)+
\frac{1}{3}k_1^2 k_2^4 (-\cos\Omega \sin^2\psi_2-
\end{displaymath}
\begin{displaymath}
\cos\Omega \cos^2\psi_2+\cos\Omega+ \cos\psi_1 \cos^3\psi_2-
\cos\psi_1 \cos\psi_2+ \cos\psi_1 \sin^2\psi_2 \cos\psi_2),
\end{displaymath}
\begin{equation}
\label{f30}
N^{(3)}_{PS}=
a_2^2 k_1^2 k_2^2 \Bigl(\frac{2}{3} \cos\Omega^2 \cos\psi_1 \cos\psi_2-
\frac{4}{3} \cos\Omega \cos^2\psi_1 \cos^2\psi_2-
\frac{2}{3} \cos\Omega \cos^2\psi_1+
\end{equation}
\begin{displaymath}
\frac{2}{3} \cos^3\psi_1 \cos^3\psi_2-
\frac{14}{3} \cos^3\psi_1 \cos\psi_2+
2 \cos\psi_1 \cos\psi_2+
\frac{2}{3} \cos^3\psi_1 \sin^2\psi_2 \cos\psi_2-
\end{displaymath}
\begin{displaymath}
\frac{2}{3} \sin^2\psi_1 \cos\psi_1 \cos\psi_2-
\frac{2}{3} \cos\psi_1 \sin^2\psi_2 \cos\psi_2\Bigr)+
a_2^2 k_1^3 k_2 \Bigl(-\frac{1}{3}\cos\Omega\cos^3\psi_1 \cos\psi_2+
\end{displaymath}
\begin{displaymath}
\cos\Omega \cos\psi_1 \cos\psi_2-\frac{1}{3}\cos\Omega \sin^2\psi_1 \cos\psi_1 \cos\psi_2-
\cos^2\psi_1 \cos^2\psi_2+\frac{1}{3}\cos^4\psi_1\cos^2\psi_2+
\end{displaymath}
\begin{displaymath}
\frac{1}{3}\sin^2\psi_1\cos^2\psi_1\cos^2\psi_2\Bigr)+
k_1^4 k_2^2 \Bigl(-\frac{4}{3}\cos\Omega^2\cos\psi_1 
\cos\psi_2-\frac{2}{3}\cos\Omega \cos^2\psi_1 \cos^2\psi_2-
\end{displaymath}
\begin{displaymath}
\frac{2}{3} \cos\Omega \sin^2\psi_1 \cos^2\psi_2-
\frac{1}{3} \cos\Omega\sin^2\psi_1-\frac{1}{3}\cos\Omega \cos^2\psi_1+2\cos\Omega \cos^2\psi_2+
\frac{1}{3}\cos\Omega+
\end{displaymath}
\begin{displaymath}
\frac{1}{3} \cos^3\psi_1 \cos\psi_2-
\frac{1}{3} \cos\psi_1 \cos\psi_2+
\frac{1}{3} \sin^2\psi_1 \cos\psi_1 \cos\psi_2\Bigr)+
k_1^3 k_2^3 \Bigl(\frac{4}{3} \cos\Omega^2 \cos^2\psi_1+
\frac{2}{3} \cos^2\Omega+
\end{displaymath}
\begin{displaymath}
\frac{2}{3}\cos\Omega\cos^3\psi_1 \cos\psi_2-\frac{10}{3} \cos\Omega \cos\psi_1 \cos\psi_2+
\frac{2}{3} \cos\Omega \sin^2\psi_1 \cos\psi_1 \cos\psi_2+\frac{2}{3} \cos^2\psi_1 \cos^2\psi_2+
\end{displaymath}
\begin{displaymath}
\frac{2}{3} \cos^2\psi_1 \sin^2\psi_2-
\frac{2}{3} \sin^2\psi_1-2 \cos^2\psi_1-\frac{2}{3} \sin^2\psi_2+2\Bigr).
\end{displaymath}

Similar amplitudes in the case of scalar mesons have the form:
\begin{equation}
\label{f31}
T^{(1)}_{S}=\frac{Z^3\alpha^4}{16\pi^4 F_\pi^2}
\int\frac{d^4k_1}{\pi^2}
\int\frac{d^4k_2}{\pi^2}
[g^{\mu\nu}tk_1-t^\mu k_1^\nu]
[g^{\sigma\lambda}k_2(-k_2-k_1-t)-
k_2^\lambda(-k_2^\sigma-k_1^\sigma-t^\sigma)]
\times
\end{equation}
\begin{displaymath}
Tr\Biggl\{
\hat\Pi^\ast_{S}
\frac{(\hat q_1+m_1)}
{2m_1}\gamma^\mu\frac{(\hat p_1+m_1)}{2m_1}
\hat\Pi_{S}
\frac{(\hat p_2-m_2)}{2m_2}\gamma^\nu
\frac{(\hat k_1-\hat p_2+m_2)}{(k_1-p_2)^2-m_2^2}\gamma^\sigma\times
\end{displaymath}
\begin{displaymath}
\frac{(\hat q_2+\hat k_2+m_2)}{(q_2+k_2)^2-m_2^2}\gamma^\lambda
\frac{(\hat q_2-m_2)}{2m_2}\Biggr\}\frac{F_{Ps}(0,k_1^2)F_{Ps}((k_1+k_2)^2,k_2^2)}{t^2k_1^2k_2^2(k_1+k_2)^2\left(
k_1^2+\frac{M_{Ps}^2}{\Lambda^2}\right)},
\end{displaymath}
\begin{equation}
\label{f32}
T^{(2)}_{S}=\frac{Z^3\alpha^4}{16\pi^4 F_\pi^2}
\int\frac{d^4k_1}{\pi^2}
\int\frac{d^4k_2}{\pi^2}
[g^{\mu\nu}tk_1-t^\mu k_1^\nu]
[g^{\sigma\lambda}k_2(-k_2-k_1-t)-
k_2^\lambda(-k_2^\sigma-k_1^\sigma-t^\sigma)]
\times
\end{equation}
\begin{displaymath}
Tr\Biggl\{
\hat\Pi^\ast_{S}\frac{(\hat q_1+m_1)}
{2m_1}\gamma^\mu\frac{(\hat p_1+m_1)}{2m_1}
\hat\Pi_{S}\frac{(\hat p_2-m_2)}{2m_2}\gamma^\sigma
\frac{(\hat k_1+\hat q_2+m_2)}{(k_1+q_2)^2-m_2^2}\gamma^\nu\times
\end{displaymath}
\begin{displaymath}
\frac{(\hat q_2-\hat k_2+m_2)}{(q_2-k_2)^2-m_2^2}\gamma^\lambda
\frac{(\hat q_2-m_2)}{2m_2}\Biggr\}\frac{F_{Ps}(0,k_1^2)F_{Ps}((k_1+k_2)^2,k_2^2)}{t^2k_1^2k_2^2(k_1+k_2)^2\left(
(k_1+k_2)^2+\frac{M_{Ps}^2}{\Lambda^2}\right)},
\end{displaymath}
\begin{equation}
\label{f33}
T^{(3)}_{S}=\frac{Z^3\alpha^4}{16\pi^4 F_\pi^2}
\int\frac{d^4k_1}{\pi^2}
\int\frac{d^4k_2}{\pi^2}
[g^{\mu\nu}tk_1-t^\mu k_1^\nu]
[g^{\sigma\lambda}k_2(-k_2-k_1-t)-
k_2^\lambda(-k_2^\sigma-k_1^\sigma-t^\sigma)]
\times
\end{equation}
\begin{displaymath}
Tr\Biggl\{
\hat\Pi^\ast_{S}\frac{(\hat q_1+m_1)}
{2m_1}\gamma^\mu\frac{(\hat p_1+m_1)}{2m_1}\hat\Pi_{S}\frac{(\hat p_2-m_2)}{2m_2}\gamma^\sigma
\frac{(\hat k_2-\hat p_2+m_2)}{(k_2-p_2)^2-m_2^2}\gamma^\sigma\times
\end{displaymath}
\begin{displaymath}
\frac{(-\hat q_2-\hat k_1+m_2)}{(q_2+k_2)^2-m_2^2}\gamma^\nu
\frac{(\hat q_2-m_2)}{2m_2}\Biggr\}\frac{F_{Ps}(0,k_1^2)F_{Ps}((k_1+k_2)^2,k_2^2)}{t^2k_1^2k_2^2(k_1+k_2)^2\left(
k_1^2+\frac{M_{Ps}^2}{\Lambda^2}\right)}.
\end{displaymath}

The general structure of the expressions in the numerators of the amplitudes \eqref{f31}, \eqref{f32}, \eqref{f33} is completely analogous to \eqref{f28}, \eqref{f29}, \eqref{f30}.
Omitting the intermediate expressions, we present in 
Appendix~\ref{app1} their explicit expressions and in 
Table~\ref{tb1} the obtained numerical results of the various contributions using the same parametrization of the transition form factors as in the previous sections.

We also calculated the contribution of axial vector mesons in three-photon interactions of one lepton.
To estimate the contribution of this type, the meson propagator is chosen in the unitary gauge:
\begin{equation}
\label{f34}
D^{\mu\nu}(k)=\frac{1}{(k^2-M_A^2)}\left(g^{\mu\nu}-\frac{k^\mu k^\nu}{M_A^2}\right).
\end{equation}

The amplitudes of interaction processes with such mesons are also presented in integral form:
\begin{equation}
\label{f35}
T^{(1)}_{AV}=\frac{Z^3\alpha^4}{16\pi^4 F_\pi^2}
\int\frac{d^4k_1}{\pi^2}
\int\frac{d^4k_2}{\pi^2}
\Gamma^{\mu\nu\sigma\lambda}_1
\frac{F_{AV}(0,k_1^2)F_{AV}((k_1+k_2)^2,k_2^2)}{t^2k_1^2k_2^2(k_1+k_2)^2\left(
k_1^2+\frac{M_{Ps}^2}{\Lambda^2}\right)}
\times
\end{equation}
\begin{displaymath}
Tr\Bigl\{
\hat\Pi^\ast_S
\frac{(\hat q_1+m_1)}
{2m_1}\gamma^\mu\frac{(\hat p_1+m_1)}{2m_1}
\hat\Pi_S
\frac{(\hat p_2-m_2)}{2m_2}\gamma^\nu
\frac{(\hat k_1-\hat p_2+m_2)}{(k_1-p_2)^2-m_2^2}\gamma^\sigma
\frac{(\hat q_2+\hat k_2+m_2)}{(q_2+k_2)^2-m_2^2}\gamma^\lambda
\frac{(\hat q_2-m_2)}{2m_2}\Bigr\},
\end{displaymath}
\begin{equation}
\label{f36}
T^{(2)}_{AV}=\frac{Z^3\alpha^4}{16\pi^4 F_\pi^2}
\int\frac{d^4k_1}{\pi^2}
\int\frac{d^4k_2}{\pi^2}
\Gamma^{\mu\nu\sigma\lambda}_1
\frac{F_{AV}(0,k_1^2)F_{AV}((k_1+k_2)^2,k_2^2)}{t^2k_1^2k_2^2(k_1+k_2)^2\left(
k_1^2+\frac{M_{Ps}^2}{\Lambda^2}\right)}
\times
\end{equation}
\begin{displaymath}
Tr\Bigl\{
\hat\Pi^\ast_S\frac{(\hat q_1+m_1)}
{2m_1}\gamma^\mu\frac{(\hat p_1+m_1)}{2m_1}
\hat\Pi_S\frac{(\hat p_2-m_2)}{2m_2}\gamma^\sigma
\frac{(\hat k_1+\hat q_2+m_2)}{(k_1+q_2)^2-m_2^2}\gamma^\nu
\frac{(\hat q_2-\hat k_2+m_2)}{(q_2-k_2)^2-m_2^2}\gamma^\lambda
\frac{(\hat q_2-m_2)}{2m_2}\Bigr\},
\end{displaymath}
\begin{equation}
\label{f37}
T^{(3)}_{AV}=\frac{Z^3\alpha^4}{16\pi^4 F_\pi^2}
\int\frac{d^4k_1}{\pi^2}
\int\frac{d^4k_2}{\pi^2}
\Gamma^{\mu\nu\sigma\lambda}_2
\frac{F_{AV}(0,k_1^2)F_{AV}((k_1+k_2)^2,k_2^2)}{t^2k_1^2k_2^2(k_1+k_2)^2\left(
k_1^2+\frac{M_{Ps}^2}{\Lambda^2}\right)}\times
\end{equation}
\begin{displaymath}
Tr\Bigl\{
\hat\Pi^\ast_S\frac{(\hat q_1+m_1)}
{2m_1}\gamma^\mu\frac{(\hat p_1+m_1)}{2m_1}\hat\Pi_S\frac{(\hat p_2-m_2)}{2m_2}\gamma^\sigma
\frac{(\hat k_2-\hat p_2+m_2)}{(k_2-p_2)^2-m_2^2}\gamma^\sigma
\frac{(-\hat q_2-\hat k_1+m_2)}{(q_2+k_2)^2-m_2^2}\gamma^\nu
\frac{(\hat q_2-m_2)}{2m_2}\Bigr\},
\end{displaymath}
where the tensors determining the hadronic scattering 
of light by light are equal to
\begin{equation}
\label{f38}
\Gamma^{\mu\nu\sigma\lambda}_{1}=
\varepsilon^{\mu\nu\alpha\beta}(t^\alpha k_1^2-k_1^\alpha t^2)
\varepsilon^{\sigma\lambda\rho\omega}\Bigl[-
(k_1+k_2+t)^\rho k_2^2-k_2^\rho(k_1+k_2+t)^2\Bigr],~
\Gamma^{\mu\nu\sigma\lambda}_{3}=-\Gamma^{\mu\nu\sigma\lambda}_{1},
\end{equation}
\begin{equation}
\label{f39}
\Gamma^{\mu\nu\sigma\lambda}_{2}=
\varepsilon^{\mu\nu\alpha\beta}(t^\alpha k_1^2-k_1^\alpha t^2)
\varepsilon^{\sigma\lambda\rho\omega}\Bigl[
k_2^\rho (k_1+k_2+t)^2+(k_1^\rho+k_2^\rho+t^\rho)k_2^2\Bigr].
\end{equation}

Then the numerators of three amplitudes with the axial vector meson have the form shown in Appendix~\ref{app2}.

\section{Conclusion}
\label{concl}

In this work we continue the study \cite{apm2023} the hadronic contribution of light-by-light scattering to the energy interval (1S-2S) in muonium which is of the seventh order in $\alpha$.
All corrections in the muonium energy spectrum considered above are determined by such amplitudes of particle interaction in which the production of mesons occurs as a result of the fusion of two photons.
The contributions of pseudoscalar, scalar and axial vector mesons are taken into account.
The numerical values of the contributions of individual mesons and the total contribution are presented in the Table~\ref{tb1}.

The key element in constructing hadronic interaction amplitudes is the form factor of the transition of two virtual photons into a meson.
Information on such form factors is extracted from various reactions for different mesons. There are many collaborations that study the form factors of the transition of two photons to mesons and systematically refine their results \cite{aihara,hamm,hamm1,L3C,tff1,tff2,tff3,tff4}. In our calculations of the hadronic contribution of
light-by-light scattering, we use a number of known parameterizations for various mesons, the calculation results with which are in agreement within the error limits. The simplest and most convenient parameterization, which allows numerical calculations of the contribution based on the obtained integral relations \eqref{f2}, \eqref{f15}, \eqref{f18}, \eqref{f24}, is the parameterization of the vector dominance model.

The calculation results are presented in final Table~\ref{tb1}. The series of values in Table~\ref{tb1} is 0, which means that this contribution has a higher order in $\alpha$ than the contribution of order $\alpha^7$ considered by us.
The calculation results also show that the structure of the constructed hadronic interaction amplitudes is such that all contributions of axial vector mesons are very small.
It is possible to single out only the contributions of scalar and pseudoscalar mesons with the structure of three-photon interaction of one lepton (see Fig.~\ref{ris2}), the contribution of which is decisive for obtaining the total hadronic contribution for the interval $(1S-2S)$.
It is necessary to keep in mind that in addition to the three amplitudes shown in Fig.~\ref{ris2}, three more similar amplitudes with photons interacting with another lepton are also taken into account.

\begin{figure}[htbp]
\includegraphics[width=0.4\textwidth]{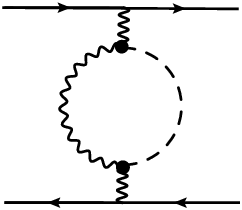}
\caption{Hadronic light-by-light scattering amplitudes in one photon interaction. Wavy line corresponds to the virtual photon. 
Dashed line corresponds to the virtual meson.
The bold dot denotes the form factor of the transition of two photons into a meson.}
\label{ris3}
\end{figure}

In conclusion, let us consider the estimate of the contribution of another hadronic amplitude, which was discussed in works \cite{acz1,acz2}, but was not considered by us previously. 
This amplitude has the form of an insertion of a polarization operator with a pion and a photon in an intermediate state in a single-photon interaction.
The renormalized polarization tensor has the form:
\begin{equation}
\label{f40}
P^{\mu\nu}_R(k^2)=(g^{\mu\nu} k^2-k^\mu k^\nu)P_R(k^2).
\end{equation}
The invariant function $P_R(k^2)$ was presented in \cite{acz1} in integral form (see Eq.(13) for $J(k^2)$ \cite{acz1}). 
For our purposes it is sufficient to have the value $J(0)$, which is obtained after integration over the Feynman parameter $x$ as follows:
\begin{equation}
\label{f41}
P_R(k^2)=\frac{\alpha^2}{16\pi^2 F_\pi^2}k^2 J(0),
\end{equation}
\begin{displaymath}
J(0)=-1+\frac{7\varepsilon^4-8\varepsilon^2-(6\varepsilon^4+8\varepsilon^2-2)\ln\varepsilon+1}
{6(1-\varepsilon^2)^4},
\end{displaymath}
where $\varepsilon=m_\pi/M_\rho$.
As a result, the shift of the S-states is determined by the following formula in the case of pseudoscalar mesons:
\begin{equation}
\label{f42}
\Delta E_{Ps}(nS)=-\frac{\mu^3\alpha^6}{24\pi^4F_\pi^2 n^3}(5-2\varepsilon),
\end{equation}
where we make an expansion of a function \eqref{f41} in parameter $\varepsilon$ omitting terms of second order.
 For the numerical contribution of $\pi$-meson to the energy interval in muonium we then obtain the value
$\Delta E_\pi(1S-2S)=-1.79~Hz$.

In a similar way, one can find the contribution of the scalar meson to both the polarization operator and the energy interval in the form:
\begin{equation}
\label{f43}
\Delta E_{S}(nS)=-\frac{2\mu^3\alpha^6 A_S^2}{9n^3}
[35+4\varepsilon_S+144\varepsilon^2_S\ln\varepsilon_S],
\end{equation}
where $\varepsilon_S=M_S/\Lambda_S$.
Numerically, the contribution of $\sigma$ meson
even slightly exceeds the previous one for $\pi$ meson: $\Delta E_\sigma(1S-2S)=-9.53~Hz$.

If the measurement accuracy of the (1S-2S) transition in muonium is increased, it will be possible to obtain the electron-to-muon mass ratio with a higher accuracy than that currently known:
$m_\mu/m_e=206.7682827(46)(22~ppb)$ \cite{codata2024}.
The theoretical value of the transition frequency (1S-2S) 2455528934.9 (0.3) MHz
obtained in the work \cite{egs2001} agrees with the experimental result \eqref{f0} within the error limits.
Despite the estimate of possible value of hadronic contribution of light-by-light scattering of order $\alpha^7$ to the energy spectrum of S-states given at the beginning, real calculations show that the contribution of the amplitudes in Fig.~\ref{ris1}-\ref{ris2} is small and still significantly less than the new accuracy of measuring the energy interval in muonium, which is planned in the Mu-MASS experiment \cite{crivelli1,ben}. More significant in magnitude 
due to a decrease in the $\alpha$ degree
are the contributions of the amplitudes of single-photon interaction in Fig.~\ref{ris3} with the production of different mesons, although they still do not reach the experimental accuracy.

\begin{acknowledgments}
The authors are grateful to A. E. Radzhabov and A. S.
Zhevlakov for useful discussions. This work is supported
by Russian Science Foundation (Grant No. RSF 23-
22-00143).
\end{acknowledgments}

\appendix
\section{Structural elements of interaction amplitudes with scalar mesons.}
\label{app1} 
\begin{equation}
\label{pr1}
N^{(1)}_S   =
k_1^2k_2^4  ( \cos\psi_1\cos\psi_2 - \cos\psi_1\cos\psi_2\sin^2\psi_1 - \cos^3\psi_1\cos\psi_2 - \cos\Omega - 2\cos\Omega\cos\psi_1^2 )
\end{equation}
\begin{displaymath}
+ k_1^3k_2^3  ( \frac{2}{3}\sin^2\psi_1 + \frac{2}{3}\cos^2\psi_2\sin^2\psi_1 + 2\cos\psi_1^2 + \frac{4}{3}
\cos\psi_1^2\sin^2\psi_2 + 2\cos\psi_1^2\cos^2\psi_2 -
\end{displaymath}
\begin{displaymath}
\frac{2}{3}\cos\Omega\cos\psi_1\cos\psi_2 - \frac{2}{3}\cos\Omega\cos\psi_1\cos\psi_2\sin^2\psi_1 - 
\frac{2}{3}\cos\Omega\cos^3\psi_1\cos\psi_2 +
\end{displaymath}
\begin{displaymath}
\frac{2}{3}\cos^2\Omega - \frac{4}{3}\cos^2\Omega\cos\psi_1^2 )
+ k_1^4k_2^2  ( \frac{5}{3}\cos\psi_1\cos\psi_2 - \cos\psi_1\cos\psi_2\sin^2\psi_1 - \cos^3\psi_1\cos\psi_2 + 
\end{displaymath}
\begin{displaymath}
\frac{1}{3}
\cos\Omega + \cos\Omega\sin^2\psi_1 + \frac{2}{3}\cos\Omega\cos^2\psi_2\sin^2\psi_1 + \cos\Omega\cos\psi_1^2 + 
\frac{2}{3}\cos\Omega\cos\psi_1^2\cos^2\psi_2 + 
\end{displaymath}
\begin{displaymath}
\frac{4}{3}\cos^2\Omega\cos\psi_1\cos\psi_2 )
+ a_2^2k_1k_2^3  (  - \frac{2}{3}\cos\psi_1^2\cos^2\psi_2 + \frac{2}{3}\cos\Omega\cos\psi_1\cos\psi_2 )+
\end{displaymath}
\begin{displaymath}
a_2^2k_1^2k_2^2  (  - \frac{2}{3}\cos\psi_1\cos\psi_2 + \frac{2}{3}\cos\psi_1\cos\psi_2\sin^2\psi_1 + \frac{5}{3}
\cos\psi_1\cos^3\psi_2 - \frac{1}{3}\cos\psi_1\cos^3\psi_2\sin^2\psi_1 +
\end{displaymath}
\begin{displaymath}
2\cos^3\psi_1\cos\psi_2 + \frac{4}{3}\cos^3\psi_1
\cos\psi_2\sin^2\psi_2 + \cos^3\psi_1\cos^3\psi_2 + \frac{2}{3}\cos\Omega - \frac{5}{3}\cos\Omega\cos^2\psi_2 - 
\end{displaymath}
\begin{displaymath}
\frac{8}{3}
\cos\Omega\cos\psi_1^2\cos^2\psi_2 + \frac{2}{3}\cos^2\Omega\cos\psi_1\cos\psi_2 )
+ a_2^2k_1^3k_2  ( \frac{5}{3}\cos\psi_1^2\cos^2\psi_2 - \cos\psi_1^2\cos^2\psi_2\sin^2\psi_1 - 
\end{displaymath}
\begin{displaymath}
\cos^4\psi_1\cos^2\psi_2 + \frac{1}{3}\cos\Omega\cos\psi_1\cos\psi_2 + \cos\Omega\cos\psi_1\cos\psi_2\sin^2\psi_1 + \cos\Omega\cos^3\psi_1\cos\psi_2 )+ 
\end{displaymath}
\begin{displaymath}
a_2^4k_1k_2  (  - \frac{2}{3}\cos\psi_1^2\cos^2\psi_2 + \frac{2}{3}\cos\Omega\cos\psi_1\cos\psi_2 ),
\end{displaymath}
\begin{equation}
\label{pr2}
N^{(2)}_S=
k_1^2k_2^4(- \frac{7}{3}\cos\psi_1\cos\psi_2 - \frac{1}{3}\cos\psi_1\cos\psi_2\sin^2\psi_2 + 
\frac{2}{3}\cos\psi_1\cos\psi_2\sin^2\psi_1 - 
\end{equation}
\begin{displaymath}
\frac{1}{3}\cos\psi_1\cos^3\psi_2 + \frac{2}{3}\cos^3\psi_1\cos\psi_2 - 
\frac{2}{3}\cos\Omega + \frac{1}{3}\cos\Omega
\sin^2\psi_2 + \frac{1}{3}\cos\Omega\cos^2\psi_2 - 
\frac{2}{3}\cos\Omega\cos\psi_1^2 )+ 
\end{displaymath}
\begin{displaymath}
k_1^3k_2^3  (  - \frac{2}{3}\sin^2\psi_1 + \cos^2\psi_2 + \frac{5}{3}\cos^2\psi_2\sin^2\psi_1 - 3\cos\psi_1^2+ 
\frac{1}{3}\cos\psi_1^2\sin^2\psi_2 + 2\cos\psi_1^2\cos^2\psi_2 )+ 
\end{displaymath}
\begin{displaymath}
k_1^4k_2^2  ( \cos\psi_1\cos\psi_2 - \frac{2}{3}\cos\psi_1\cos\psi_2\sin^2\psi_2 + 
\cos\psi_1\cos\psi_2\sin^2\psi_1
- \frac{2}{3}\cos\psi_1\cos^3\psi_2 + \cos^3\psi_1\cos\psi_2- 
\end{displaymath}
\begin{displaymath}
\frac{1}{3}\cos\Omega\sin^2\psi_1 + \frac{2}{3}\cos\Omega\cos^2\psi_2+ \cos\Omega\cos\psi_1^2 )
+ a_2^2k_2^4  ( \frac{1}{3}\cos\psi_1\cos\psi_2\sin^2\psi_2 + \cos\psi_1\cos^3\psi_2 )+ 
\end{displaymath}
\begin{displaymath}
a_2^2k_1k_2^3  (  - \frac{1}{3}\sin^2\psi_2 - \frac{1}{3}\cos^2\psi_2\sin^2\psi_2 - \cos^4\psi_2 + \frac{1}{3}
 \cos\psi_1^2\sin^2\psi_2 + \frac{8}{3}\cos\psi_1^2\cos^2\psi_2- 
 \end{displaymath}
\begin{displaymath}
 \frac{1}{3}\cos\psi_1^2\cos^2\psi_2\sin^2\psi_2 - 
 \frac{1}{3}
\cos\psi_1^2\cos^4\psi_2 - \frac{11}{3}\cos\Omega\cos\psi_1\cos\psi_2 + \frac{1}{3}\cos\Omega\cos\psi_1\cos\psi_2\sin^2\psi_2 + 
\end{displaymath}
\begin{displaymath}
\frac{1}{3}
\cos\Omega\cos\psi_1\cos^3\psi_2 )
+ a_2^2k_1^2k_2^2  ( \frac{1}{3}\cos\psi_1\cos\psi_2 -\frac{1}{3}\cos\psi_1\cos\psi_2\sin^2\psi_2 -
\frac{4}{3}\cos\psi_1\cos\psi_2\sin^2\psi_1 - 
\end{displaymath}
\begin{displaymath}
4\cos\psi_1\cos^3\psi_2 + \frac{7}{3}\cos\psi_1\cos^3\psi_2\sin^2\psi_1 +\frac{1}{3}\cos^3\psi_1
\cos\psi_2 - \frac{1}{3}\cos^3\psi_1\cos\psi_2\sin^2\psi_2 +
\end{displaymath}
\begin{displaymath}
2\cos^3\psi_1\cos^3\psi_2 - \frac{1}{3}\cos\Omega + 3\cos\Omega
\cos^2\psi_2 - \frac{7}{3}\cos\Omega\cos\psi_1^2)
+ a_2^2k_1^3k_2  ( \frac{2}{3}\cos^2\psi_2\sin^2\psi_1 - 
\end{displaymath}
\begin{displaymath}
\frac{7}{3}\cos\psi_1^2\cos^2\psi_2 + \cos\psi_1^2
\cos^2\psi_2\sin^2\psi_1 + \cos^4\psi_1\cos^2\psi_2 + \frac{7}{3}\cos\Omega\cos\psi_1\cos\psi_2 - 
\end{displaymath}
\begin{displaymath}
\frac{1}{3}\cos\Omega\cos\psi_1
\cos\psi_2\sin^2\psi_1 + \cos\Omega\cos^3\psi_1\cos\psi_2 )
- \frac{1}{3}a_2^4k_2^2 \cos\psi_1\cos\psi_2\sin^2\psi_2 + 
\end{displaymath}
\begin{displaymath}
a_2^4k_1k_2  ( \frac{1}{3}\cos\psi_1^2\cos^2\psi_2 - \frac{1}{3}\cos\Omega\cos\psi_1\cos\psi_2 ),
\end{displaymath}
\begin{equation}
\label{pr3}
 N^{(3)}_S =k_1^2k_2^4  ( 7\cos\psi_1\cos\psi_2 - \cos\psi_1\cos\psi_2\sin^2\psi_1 - \cos^3\psi_1\cos\psi_2 - \cos\Omega
- 2\cos\Omega\cos\psi_1^2 )
\end{equation}
\begin{displaymath}
+ k_1^3k_2^3  (- \frac{1}{3}\sin^2\psi_1 + \frac{2}{3}\cos^2\psi_2\sin^2\psi_1 + \cos\psi_1^2 - \frac{2}{3}
\cos\psi_1^2\sin^2\psi_2 + \frac{22}{3}\cos\Omega\cos\psi_1\cos\psi_2 - 
\end{displaymath}
\begin{displaymath}
\frac{2}{3}\cos\Omega\cos\psi_1\cos\psi_2\sin^2\psi_1 - \frac{2}{3}\cos\Omega\cos^3\psi_1\cos\psi_2 - 
\frac{4}{3}\cos^2\Omega - \frac{4}{3}\cos^2\Omega\cos\psi_1^2 )+ 
\end{displaymath}
\begin{displaymath}
k_1^4k_2^2  ( \cos\psi_1\cos\psi_2 - \frac{1}{3}\cos\psi_1\cos\psi_2\sin^2\psi_1 -\frac{1}{3}\cos^3\psi_1\cos\psi_2 + 
\frac{1}{3}\cos\Omega\sin^2\psi_1 + 
\end{displaymath}
\begin{displaymath}
\frac{2}{3}\cos\Omega\cos^2\psi_2\sin^2\psi_1 + 
\frac{1}{3}\cos\Omega\cos\psi_1^2 + \frac{2}{3}\cos\Omega
\cos\psi_1^2\cos^2\psi_2 + \frac{4}{3}\cos^2\Omega\cos\psi_1\cos\psi_2 )+ 
\end{displaymath}
\begin{displaymath}
a_2^2k_1k_2^3  ( 2\cos\psi_1^2 + \frac{16}{3}\cos\psi_1^2\cos^2\psi_2 - \frac{4}{3}\cos\Omega\cos\psi_1\cos\psi_2 )
+ a_2^2k_1^2k_2^2  (  -\frac{7}{3}\cos\psi_1\cos\psi_2 - 
\end{displaymath}
\begin{displaymath}
\frac{1}{3}\cos\psi_1\cos\psi_2\sin^2\psi_1 + \frac{5}{3}
\cos\psi_1\cos^3\psi_2 - \frac{1}{3}\cos\psi_1\cos^3\psi_2\sin^2\psi_1 - \frac{1}{3}\cos^3\psi_1\cos\psi_2 - 
\end{displaymath}
\begin{displaymath}
\frac{2}{3}\cos^3\psi_1\cos\psi_2\sin^2\psi_2 - \cos^3\psi_1\cos^3\psi_2 + \frac{1}{3}\cos\Omega + 
\frac{1}{3}\cos\Omega\cos^2\psi_2 + \frac{7}{3}
\cos\Omega\cos\psi_1^2 + 
\end{displaymath}
\begin{displaymath}
\frac{16}{3}\cos\Omega\cos\psi_1^2\cos^2\psi_2 - 
\frac{4}{3}\cos^2\Omega\cos\psi_1\cos\psi_2 )
+ a_2^2k_1^3k_2  ( \frac{7}{3}\cos\psi_1^2\cos^2\psi_2 - 
\end{displaymath}
\begin{displaymath}
\frac{1}{3}\cos\psi_1^2\cos^2\psi_2\sin^2\psi_1 - 
\frac{1}{3}\cos^4\psi_1\cos^2\psi_2 - \frac{7}{3}\cos\Omega\cos\psi_1\cos\psi_2 + 
\frac{1}{3}\cos\Omega\cos\psi_1\cos\psi_2\sin^2\psi_1  
\end{displaymath}
\begin{displaymath}
+\frac{1}{3}\cos\Omega\cos^3\psi_1\cos\psi_2 )+ 
a_2^4k_1k_2  (  - \frac{1}{3}\cos\psi_1^2\cos^2\psi_2 + \frac{1}{3}\cos\Omega\cos\psi_1\cos\psi_2 ).
\end{displaymath}
\section{Structural elements of interaction amplitudes with axial vector mesons.}
\label{app2} 
\begin{equation}
\label{ff4}
N^{(1)}_{AV}=
k_1^3k_2^5  ( 4 + 4\cos^2\psi_2 + \frac{4}{3}\cos^2\psi_1\sin^2\psi_2 + \frac{4}{3}\cos^2\psi_1\cos^2\psi_2
 - \frac{16}{3}\cos\Omega\cos\psi_1\cos\psi_2 + 
 \end{equation}
 \begin{displaymath}
 4\cos\Omega^2 )
+ k_1^4k_2^4  \Bigl( 8\cos\psi_1\cos\psi_2 + 4\cos\Omega - \frac{4}{3}\cos\Omega\sin^2\psi_2 + 
\frac{4}{3}\cos\Omega
\cos^2\psi_1\sin^2\psi_2 +
\end{displaymath}
\begin{displaymath}
\frac{4}{3}\cos\Omega\cos^2\psi_1\cos^2\psi_2 - \frac{16}{3}\cos\Omega^2\cos\psi_1\cos\psi_2 + 4
\cos\Omega^3 \Bigr)
+ k_1^5k_2^3  \Bigl( 3 - \frac{7}{3}\sin^2\psi_1 + 
\end{displaymath}
\begin{displaymath}
2\cos^2\psi_2 - \frac{4}{3}\cos^2\psi_2\sin^2\psi_1 + 3
\cos^2\psi_1 + \frac{2}{3}\cos^2\psi_1\sin^2\psi_2 -
\frac{2}{3}\cos^2\psi_1\cos^2\psi_2 - 
\end{displaymath}
\begin{displaymath}
\frac{4}{3}\cos\Omega\cos\psi_1\cos\psi_2
\sin^2\psi_2 - \frac{4}{3}\cos\Omega\cos\psi_1\cos\psi_2^3 + \frac{2}{3}\cos\Omega^2 - \frac{8}{3}\cos\Omega^2\cos^2\psi_2 \Bigr)+
\end{displaymath}
\begin{displaymath}
k_1^6k_2^2  \Bigl( 2\cos\psi_1\cos\psi_2 - \frac{2}{3}\cos\psi_1\cos\psi_2\sin^2\psi_2 - \frac{2}{3}\cos\psi_1\cos\psi_2^3
- \frac{4}{3}\cos\Omega\cos^2\psi_2 \Bigr)+ 
\end{displaymath}
\begin{displaymath}
a_2^2 k_1^2k_2^4 \Bigl( 4\cos\psi_1\cos\psi_2 + 4\cos\psi_1\cos\psi_2^3 - 4\cos\Omega\cos^2\psi_1
- 8\cos\Omega\cos^2\psi_1\cos^2\psi_2 + 
\end{displaymath}
\begin{displaymath}
4\cos\Omega^2\cos\psi_1\cos\psi_2 \Bigr)
+ a_2^2 k_1^3k_2^3  \Bigl( \frac{4}{3}\cos^2\psi_1\sin^2\psi_2 + 8\cos^2\psi_1\cos^2\psi_2 + \frac{4}{3}
\cos^2\psi_1\cos^2\psi_2\sin^2\psi_2 + 
\end{displaymath}
\begin{displaymath}
\frac{4}{3}\cos^2\psi_1\cos\psi_2^4 + 2\cos\psi_1^4 + 8\cos\Omega\cos\psi_1
\cos\psi_2 - \frac{4}{3}\cos\Omega\cos\psi_1\cos\psi_2\sin^2\psi_2+ 
\end{displaymath}
\begin{displaymath}
\frac{8}{3}\cos\Omega\cos\psi_1\cos\psi_2^3 - 4\cos\Omega
\cos\psi_1^3\cos\psi_2 - 4\cos\Omega^2\cos^2\psi_1 - 8\cos\Omega^2\cos^2\psi_1\cos^2\psi_2 + 
\end{displaymath}
\begin{displaymath}
4\cos\Omega^3
\cos\psi_1\cos\psi_2 \Bigr)
 + a_2^2 k_1^4k_2^2  \Bigl( 3\cos\psi_1\cos\psi_2 - \frac{4}{3}\cos\psi_1\cos\psi_2\sin^2\psi_2 - \frac{7}{3}\cos\psi_1
\cos\psi_2\sin^2\psi_1 + 
\end{displaymath}
\begin{displaymath}
2\cos\psi_1\cos\psi_2^3 - \frac{4}{3}\cos\psi_1\cos\psi_2^3\sin^2\psi_1 - \frac{1}{3}\cos\psi_1^3
\cos\psi_2 - \frac{4}{3}\cos\psi_1^3\cos\psi_2^3 - \frac{2}{3}\cos\Omega\cos^2\psi_1 + 
\end{displaymath}
\begin{displaymath}
\frac{8}{3}\cos\Omega\cos^2\psi_1\cos^2\psi_2
+ \frac{14}{3}\cos\Omega^2\cos\psi_1\cos\psi_2 \Bigr)
+ \frac{2}{3}a_2^2 k_1^5k_2  \Bigl(5\cos^2\psi_1\cos^2\psi_2 +\cos\Omega\cos\psi_1\cos\psi_2 \Bigr),
\end{displaymath}
\begin{equation}
\label{ff5}
N^{(2)}_{AV}=
k_1^3k_2^5  \Bigl( 2 + \frac{2}{3}\sin^2\psi_2 - 2\cos^2\psi_2 - \frac{8}{3}\cos^2\psi_1\sin^2\psi_2 - 
\frac{8}{3}
\cos^2\psi_1\cos^2\psi_2 - 
\end{equation}
\begin{displaymath}
\frac{16}{3}\cos\Omega\cos\psi_1\cos\psi_2 \Bigr)+
 k_1^4k_2^4  \Bigl(  - \frac{28}{3}\cos\psi_1\cos\psi_2 + 
 \frac{16}{3}\cos\psi_1\cos\psi_2\sin^2\psi_2 + 
\frac{4}{3}\cos\psi_1
\cos\psi_2\sin^2\psi_1 + 
\end{displaymath}
\begin{displaymath}
\frac{16}{3}\cos\psi_1\cos\psi_2^3 + 
\frac{4}{3}\cos\psi_1^3\cos\psi_2 + \frac{10}{3}\cos\Omega + \frac{2}{3}\cos\Omega\sin^2\psi_2 - \frac{22}{3}\cos\Omega\cos^2\psi_2 - \frac{16}{3}\cos\Omega\cos^2\psi_1 - 
\end{displaymath}
\begin{displaymath}
\frac{8}{3}\cos\Omega\cos^2\psi_1
\sin^2\psi_2 - \frac{8}{3}\cos\Omega\cos^2\psi_1\cos^2\psi_2 - \frac{16}{3}\cos\Omega^2\cos\psi_1\cos\psi_2 \Bigr)
+ k_1^5k_2^3  \Bigl( 1/3\sin^2\psi_2 + 
\end{displaymath}
\begin{displaymath}
\sin^2\psi_1 - \cos^2\psi_2 - 4\cos^2\psi_2\sin^2\psi_1 - 3
\cos^2\psi_1 - \frac{4}{3}\cos^2\psi_1\sin^2\psi_2 - \frac{16}{3}\cos^2\psi_1\cos^2\psi_2 - 
\end{displaymath}
\begin{displaymath}
\frac{20}{3}\cos\Omega\cos\psi_1\cos\psi_2
+ \frac{8}{3}\cos\Omega\cos\psi_1\cos\psi_2\sin^2\psi_2 + \frac{8}{3}\cos\Omega\cos\psi_1\cos\psi_2^3 - \frac{8}{3}\cos\Omega^2
\cos^2\psi_2 \Bigr)+
\end{displaymath}
\begin{displaymath}
k_1^6k_2^2  \Bigl(  - 4\cos\psi_1\cos\psi_2 + \frac{4}{3}\cos\psi_1\cos\psi_2\sin^2\psi_2 + \frac{4}{3}\cos\psi_1
\cos\psi_2^3 - \frac{4}{3}\cos\Omega\cos^2\psi_2 \Bigr)+
\end{displaymath}
\begin{displaymath}
a_2^2 k_1k_2^5  ( 4\cos^2\psi_1\cos^2\psi_2 - 4\cos\Omega\cos\psi_1\cos\psi_2 )+
a_2^2 k_1^2k_2^4  \Bigl( 2\cos\psi_1\cos\psi_2 + 
\end{displaymath}
\begin{displaymath}
2\cos\psi_1\cos\psi_2\sin^2\psi_2 - 6\cos\psi_1
\cos\psi_2^3 + 6\cos\psi_1^3\cos\psi_2 + 4\cos\Omega\cos^2\psi_2 - 4\cos\Omega\cos^2\psi_1 - 
\end{displaymath}
\begin{displaymath}
4\cos\Omega
\cos^2\psi_1\cos^2\psi_2 - 4\cos\Omega^2\cos\psi_1\cos\psi_2 \Bigr)
+ a_2^2 k_1^3k_2^3 \Bigl(  - \frac{4}{3}\cos^2\psi_2\sin^2\psi_2 + \frac{4}{3}\cos^2\psi_1\sin^2\psi_2 - 
\end{displaymath}
\begin{displaymath}
\frac{44}{3}
\cos^2\psi_1\cos^2\psi_2 + \frac{8}{3}\cos^2\psi_1\cos^2\psi_2\sin^2\psi_2 + \frac{8}{3}\cos^2\psi_1\cos\psi_2^4 + 
2\cos\psi_1^4 + \frac{20}{3}\cos\Omega\cos\psi_1\cos\psi_2 + 
\end{displaymath}
\begin{displaymath}
\frac{2}{3}\cos\Omega\cos\psi_1\cos\psi_2\sin^2\psi_2 - \frac{26}{3}\cos\Omega
\cos\psi_1\cos^3\psi_2 + 4\cos^2\Omega\cos^2\psi_2 - 4\cos^2\Omega\cos^2\psi_1 - 
\end{displaymath}
\begin{displaymath}
4\cos^2\Omega\cos^2\psi_1
\cos^2\psi_2 \Bigr)
+ a_2^2 k_1^4k_2^2  \Bigl(  -\cos\psi_1\cos\psi_2\sin^2\psi_2 + \cos\psi_1\cos\psi_2\sin^2\psi_1 - \frac{5}{3}
\cos\psi_1\cos\psi_2^3 - 
\end{displaymath}
\begin{displaymath}
\frac{8}{3}\cos\psi_1\cos\psi_2^3\sin^2\psi_1 - \frac{13}{3}\cos\psi_1^3\cos\psi_2 - 
\frac{8}{3}
\cos\psi_1^3\cos\psi_2^3 + \frac{2}{3}\cos\Omega\cos^2\psi_2 - \frac{2}{3}\cos\Omega\cos^2\psi_1 - 
\end{displaymath}
\begin{displaymath}
\frac{40}{3}\cos\Omega\cos^2\psi_1
\cos^2\psi_2 + 4\cos\Omega^2\cos\psi_1\cos\psi_2 \Bigr)
+ a_2^2 k_1^5k_2  \Bigl(  - \frac{14}{3}\cos^2\psi_1\cos^2\psi_2 + \frac{2}{3}\cos\Omega\cos\psi_1\cos\psi_2 \Bigr),
\end{displaymath}
\begin{equation}
\label{ff6}
N^{(3)}_{AV}=
k_1^3k_2^5  \Bigl( 4 + 4\cos^2\psi_2 - 8\cos^2\psi_1 - \frac{4}{3}\cos^2\psi_1 \sin^2\psi_2 - \frac{4}{3}
\cos^2\psi_1\cos^2\psi_2 - 
\end{equation}
\begin{displaymath}
\frac{32}{3}\cos\Omega\cos\psi_1\cos\psi_2 +
4\cos\Omega^2 \Bigr)
+ k_1^4k_2^4  \Bigl( \frac{16}{3}\cos\psi_1\cos\psi_2 + \frac{8}{3}\cos\psi_1\cos\psi_2\sin^2\psi_2 + 
\end{displaymath}
\begin{displaymath}
\frac{8}{3}\cos\psi_1
\cos\psi_2^3 + \frac{20}{3}\cos\Omega - \frac{4}{3}\cos\Omega\sin^2\psi_2 + \frac{16}{3}\cos\Omega\cos^2\psi_2 - 16\cos\Omega\cos^2\psi_1 - 
\end{displaymath}
\begin{displaymath}
\frac{4}{3}\cos\Omega\cos^2\psi_1\sin^2\psi_2 - \frac{4}{3}\cos\Omega\cos^2\psi_1\cos^2\psi_2 - \frac{32}{3}
\cos\Omega^2\cos\psi_1\cos\psi_2 + 4\cos\Omega^3 \Bigr)+
\end{displaymath}
\begin{displaymath}
k_1^5k_2^3  \Bigl( 3 - \sin^2\psi_1 + 2\cos^2\psi_2 - \frac{4}{3}\cos^2\psi_2\sin^2\psi_1 - 5\cos^2\psi_1
- \frac{2}{3}\cos^2\psi_1\sin^2\psi_2 - 2\cos^2\psi_1\cos^2\psi_2+
\end{displaymath}
\begin{displaymath}
\frac{8}{3}\cos\Omega\cos\psi_1\cos\psi_2 + \frac{4}{3}
\cos\Omega\cos\psi_1\cos\psi_2\sin^2\psi_2 + \frac{4}{3}\cos\Omega\cos\psi_1\cos\psi_2^3 + \frac{10}{3}\cos\Omega^2 + 
\end{displaymath}
\begin{displaymath}
\frac{8}{3}\cos\Omega^2\cos^2\psi_2 \Bigr)
+ k_1^6k_2^2  \Bigl( \frac{2}{3}\cos\psi_1\cos\psi_2 + \frac{2}{3}\cos\psi_1\cos\psi_2\sin^2\psi_2 + \frac{2}{3}\cos\psi_1
\cos\psi_2^3 + \frac{4}{3}\cos\Omega + 
\end{displaymath}
\begin{displaymath}
\frac{4}{3}\cos\Omega\cos^2\psi_2 \Bigr)
+ a_2^2 k_1^2k_2^4  \Bigl( 4\cos\psi_1\cos\psi_2 + 4\cos\psi_1\cos\psi_2^3 - 4\cos\Omega\cos^2\psi_1-
\end{displaymath}
\begin{displaymath}
8\cos\Omega\cos^2\psi_1\cos^2\psi_2 + 4\cos\Omega^2\cos\psi_1\cos\psi_2\Bigr )
+ a_2^2 k_1^3k_2^3  \Bigl( \frac{4}{3}\cos^2\psi_1\sin^2\psi_2 - \frac{8}{3}\cos^2\psi_1\cos^2\psi_2 +
\end{displaymath}
\begin{displaymath}
\frac{4}{3}
\cos^2\psi_1\cos^2\psi_2\sin^2\psi_2 + \frac{4}{3}\cos^2\psi_1\cos\psi_2^4 + 
2\cos\psi_1^4 + \frac{32}{3}\cos\Omega
\cos\psi_1\cos\psi_2 - 
\end{displaymath}
\begin{displaymath}
\frac{4}{3}\cos\Omega\cos\psi_1\cos\psi_2\sin^2\psi_2 + \frac{8}{3}\cos\Omega\cos\psi_1\cos\psi_2^3 - 4
\cos\Omega\cos\psi_1^3\cos\psi_2 - 4\cos\Omega^2\cos^2\psi_1 - 
\end{displaymath}
\begin{displaymath}
8\cos^2\Omega\cos^2\psi_1\cos^2\psi_2 + 
4\cos^3\Omega\cos\psi_1\cos\psi_2 \Bigr)
+ a_2^2 k_1^4k_2^2 \Bigl( 3\cos\psi_1\cos\psi_2 - \frac{4}{3}\cos\psi_1\cos\psi_2\sin^2\psi_2 - 
\end{displaymath}
\begin{displaymath}
\cos\psi_1
\cos\psi_2\sin^2\psi_1 + 2\cos\psi_1\cos\psi_2^3 - \frac{4}{3}\cos\psi_1\cos\psi_2^3\sin^2\psi_1 - 
\frac{13}{3}\cos\psi_1^3
\cos\psi_2 - 
\end{displaymath}
\begin{displaymath}
\frac{4}{3}\cos\psi_1^3\cos\psi_2^3 - \frac{2}{3}\cos\Omega\cos^2\psi_1 - 8\cos\Omega\cos^2\psi_1\cos^2\psi_2
+ \frac{22}{3}\cos\Omega^2\cos\psi_1\cos\psi_2 \Bigr)+ 
\end{displaymath}
\begin{displaymath}
a_2^2 k_1^5k_2  \Bigl(  - \cos^2\psi_1\cos^2\psi_2 + 2\cos\Omega\cos\psi_1\cos\psi_2 \Bigr).
\end{displaymath}

\bibliography{hadronic_lbl_bibliograophy}

\end{document}